\documentclass[twocolumn,fp,letterpaper]{jpsj3_arXiv}
\usepackage{txfonts}
\usepackage{bm}
\usepackage[dvipdfmx]{color}

\title{Microscopic theory of tunneling spectroscopy in Sr$_2$RuO$_4$}

\author{Keiji Yada$^1$, Alexander A. Golubov$^{2,3}$, Yukio Tanaka$^1$ and Satoshi Kashiwaya$^4$}
\inst{$^1$Department of Applied Physics, Nagoya University, Nagoya 464-8603, Japan \\
$^2$Faculty of Science and Technology and MESA+ Institute of Nanotechnology,
University of Twente, 7500 AE, Enschede, The Netherlands \\
$^3$Moscow Institute of Physics and Technology, 141700 Dolgoprudny, Moscow Region, Russia \\
$^4$National Institute of Advanced Industrial Science and Technology (AIST), Tsukuba 305-8568, Japan} 

\abst{We study the surface Andreev bound state (ABS)
of superconducting Sr$_2$RuO$_4$,
which is a candidate material for the realization of the chiral $p$-wave superconducting state.
In order to clarify the role of chiral edge modes as ABSs,
the surface density of states and the tunneling conductance
is calculated in the normal metal/Sr$_2$RuO$_4$ junction
within the framework of recursive Green's function method,
while taking into account the orbital degrees of freedom
(including Spin-Orbit interactions) with realistic material parameters.
In Sr$_2$RuO$_4$, there are two bands $\alpha$ and $\beta$ originating from
quasi-one-dimensional orbitals $d_{yz}$ and $d_{zx}$
and a two-dimensional band $\gamma$  originating from $d_{xy}$ orbital.
We discuss about the contributions of various electronic bands to LDOS
and the influence of atomic spin-orbit interaction (SOI).
In the light of our calculations,
quasi-one-dimensional model with
dominant pair potentials in $\alpha$ and $\beta$ bands
is consistent with conductance measurements in 
Au/Sr$_{2}$RuO$_{4}$ junctions.}

\begin{document}
\maketitle

\section{Introduction}\label{sec1}
Sr$_2$RuO$_4$ has attracted
much interest for its unconventional superconductivity appearing
at $T_{c} \sim 1.5$ K.\cite{Maeno94}
There have been several experimental reports
consistent with spin-triplet pairing \cite{Ishida98,Mukuda,Ishida01,Murakawa}
with broken time reversal symmetry.\cite{Luke}
The most promising candidate for the
pairing symmetry of the pair potential
is the so-called spin-triplet chiral $p$-wave state, whose pair potential
can be  represented as
$\Delta_{0}\hat{z}(k_{x} \pm i k_{y})$ in the free electron model.\cite{Maeno2}
This state is the two-dimensional analog of the $A$-phase of
superfluid $^{3}$He, 
and can be categorized as topological superconductors.\cite{Volovik,Furusaki2001}
In topological superconductors,
gapless Andreev bound states (ABSs) appear at their edges due to bulk-edge correspondence
which are symmetrically protected by the bulk energy gap.\cite{qi11,tanaka12,alicea12,Sato09,Sato10}
For the chiral $p$-wave case,
a gapless ABS with linear dispersion is generated.\cite{Matsumoto99,Furusaki2001}
Although there have been several attempts to 
observe the chiral edge modes directly,
their existence was not confirmed experimentally yet.\cite{Maeno2012,Kallin2012}

\begin{figure}[htbp]
\begin{center}
\includegraphics[width=7.5cm]{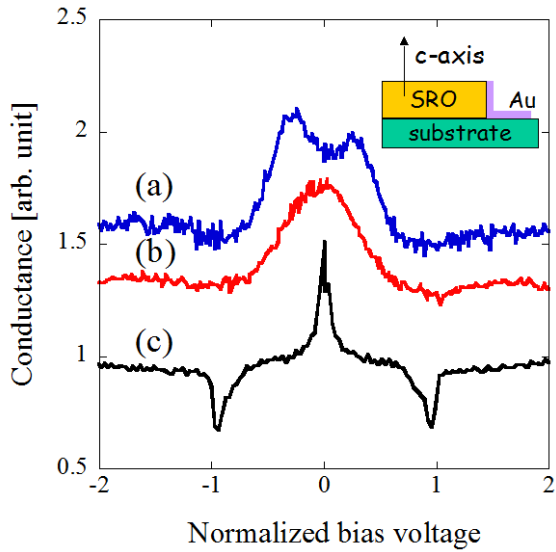}
\caption{(Color online) Conductance spectra detected using tunneling junctions formed between in-situ cleaved surface of Sr$_{2}$RuO$_{4}$ (SRO) and Au at $T\sim0.55$K.
The horizontal axes are normalized by (a),(b) 0.7 mV and (c) 0.13mV, respectively.
The vertical axes normalized by their normal-state background values and rescaled and shifted in order to clearly identify the peak structure around zero-bias level.
In the three curves, (a) and (b) are obtained at in-situ cleaved surface of SRO cited from Ref.7, and (c) is obtained at a 3K-phase surface.
The inset shows the junction structure used in ab-plane tunneling spectroscopy.}\label{fig1}
\end{center}
\end{figure}

One of the few efficient ways to detect the chiral edge modes indirectly is
tunneling conductance between normal metal/superconducting junctions.\cite{TK95,kashiwaya00}
For example, a sharp zero-bias conductance peak (ZBCP) has been predicted
in spin-singlet $d$-wave superconductors by the
ABS with flat dispersion.\cite{Hu,TK95}
Due to this flat-band dispersion, ZBCP ubiquitously emerges in actual
experiments in high $T_{\rm C}$ cuprates.
\cite{Experiment1,Experiment2,Experiment3,Experiment4,Experiment5,Experiment6,
Experiment7,Experiment8}
On the other hand, performing tunneling spectroscopy experiments in
Sr$_{2}$RuO$_{4}$ junctions is not an easy task.
Although Sr$_{2}$RuO$_{4}$ has extremely fragile surface,
tunneling spectroscopy experiments have been performed
on i) c-axis surface,\cite{Upward,Lupien,Pennec,Suderow,Firmo}
ii) in-plane,\cite{Laube,Kashiwaya11} and iii) 3K-phase.\cite{Mao,Kawamura}
As for c-axis tunneling,
several reliable data are obtained using scanning tunneling microscopy/spectroscopy (STM/S) on in-situ cleaved surfaces.
The most important feature observed in common is that conductance spectra exhibit gap feature
except for zero-bias peak obtained at the vortex cores.
The gap spectra show residual conductance at the zero-bias about
85$\%$ \cite{Upward} and 50$\%$,\cite{Lupien}
while a fully opened gap has been detected in the Al tip case.\cite{Suderow}
The variety of gap amplitudes for three different bands obtained in the
specific heat measurements \cite{Deguchi}
are not discernible for these experiments.
In the case of ab-plane tunneling spectroscopy,
due to the extreme difficulty in surface treatment,
the number of experiments reported are quite limited.
\cite{Laube,Kashiwaya11}
The data of point contact spectroscopy
and thin film junction formed on cleaved surface in vacuum,
shows the presence of the zero-bias conductance peaks in common,
which suggest the formation of gapless ABS at the in-plane edges.
Comparing with the zero-bias conductance peaks obtained in high-$T_{\rm C}$ cuprates,
the dome-like peak spreading in the whole gap amplitude observed on
Sr$_{2}$RuO$_{4}$
indicates the formation of dispersive edge states expected for chiral $p$-wave superconductors.\cite{YTK97,YTK98,Honerkamp}
The shapes of the peak show variation depending on the junction (see Fig.\ref{fig1}).
Compared to the fragile 1.5K phase surfaces,
the 3K phases of Sr$_2$RuO$_4$ tends to have stable surfaces
due to the inclusion of inert surface of Ru.
Conductance spectra obtained on these 3K phase commonly exhibit a sharp narrow peak
near zero-bias in addition to the broad peak spreading in the whole gap amplitude \cite{Mao,Kawamura}
(see curve (c) in Fig. \ref{fig1}).
However, the origin of this two-step peak has not been clarified.
One of the proposals for the superconducting state in 3K-phase is the inhomogeneous gap structure near the Ru-inclusions.\cite{Sigrist-Monien}
However, it is still an experimentally unresolved problem.
Since two-step peaks in 1.5K-phase were recently observed by one of the author,\cite{SK}
we believe that the two-step peaks do not originate from the inhomogeneous gap structure but the ABSs between Sr$_{2}$RuO$_{4}$ and a Ru-inclusion on the surface.

\begin{figure}[htbp]
\begin{center}
\includegraphics[width=6cm]{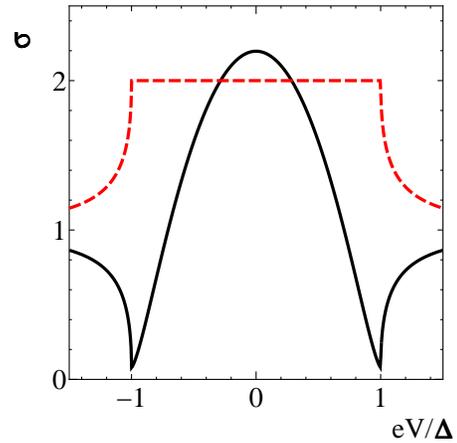}
\caption{(Color online) Normalized conductance based on effective mass approximation for high transmissivity (dashed line) and low transmissivity (solid line).}\label{fig2}
\end{center}
\end{figure}

Beside the above experimental studies,
a number of theories of tunneling spectroscopy in normal metal/spin-triplet
chiral $p$-wave superconductor junctions were formulated.\cite{YTK97,YTK98,Honerkamp,Alireza}
It was shown that the line shape of a tunneling conductance in chiral $p$-wave
superconductor junctions has a broad ZBCP (see Fig. \ref{fig2}) due to the ABS with linear
dispersion.
These theories could explain the experimental results like curve (b) in Fig. \ref{fig1}.
However, this simple picture can not give reasonable explanation for zero bias dip like curve (a) and two-step peak like curve (c) in Fig. \ref{fig1}.
While the origin of zero bias dip can be explained by mismatch of the Fermi surface \cite{Samokhin} or the anisotropy of the pair potential in momentum space,\cite{Sengupta}
there are no theories which elucidate the origin of two-step peak in conductance.
If the two-step peak in conductance comes from intrinsic nature of the superconducting Sr$_2$RuO$_4$,
then one possibility for this origin is the multi-band effect.
It is well known that this material has a two-dimensional electronic structure where
cylindrical Fermi surfaces are formed by
three bands, $\alpha$, $\beta$, and $\gamma$
sheets, which mainly originate from the Ru $4d$ orbitals.\cite{Maeno2}
The $\gamma$-band is mainly composed of two-dimensional $d_{xy}$-orbital while
$\alpha$ and $\beta$-bands are mainly composed of quasi-one-dimensional $d_{yz}$ and $d_{zx}$-orbitals.
Due to this difference of the orbital nature,
it is natural to consider that the energy gap in each orbital is different.
This difference of the energy scale might produce the two-step peak.
Actually, several microscopic mechanisms of superconductivity in Sr$_2$RuO$_4$ are proposed
based on multi-orbital model as well as single band model.\cite{Rice,Miyake,Nomura00,Kuwabara,Arita,Onari}
Some of these theories stress the superconductivity which stems from the two-dimensional $\gamma$-band,\cite{Nomura02a,Nomura02b,Yanase,Nomura05,Wang}
while there are theories where the quasi-one-dimensional $\alpha$ and $\beta$-bands were addressed.\cite{Kohmoto,Kuroki,Takimoto,Raghu}
Still, a theory of conductance considering multi-orbital effect is lacking.

In taking into account the multi-orbital effect on the conductance quantitatively,
one has to work with realistic band structures based on microscopic considerations.\cite{Devyatov}
A number of theoretical studies addressed superconductivity in the $d$-wave or $p$-wave pairing in the framework 
of tight binding model.\cite{Tanuma2002,Tanuma2001,Tanuma99,Ting00,Asano01,Imai2012,Imai2013}
However, this approach suffers from the problem of the existence of two different energy scales:
transfer integral and the pair potential.
Usually, the magnitude of transfer integral
is two to three orders of magnitude larger than that of the pair potential.
Owing to the limitations of computational resources,
one can not obtain reliable data of conductance in a finite size system
if we choose the realistic values of $\Delta$,
since, the coherence length is large and finite-size effect becomes distinct.
Since many exotic properties have been predicted in
spin-triplet $p$-wave superconductor junctions,\cite{Proximityp,Proximityp2,Proximityp3,Proximityd,Meissner3,odd1}
it is quite urgent task to calculate SDOS and
$\sigma_s$ in spin-triplet chiral $p$-wave superconductors based on a
microscopic model while taking into account electronic structures of
Sr$_{2}$RuO$_{4}$ and realistic magnitude of pair potentials.

In this paper, we calculate the SDOS and $\sigma_s$ of Sr$_{2}$RuO$_{4}$
based on the three-band model using the Green's function method for semi-infinite system.
This approach is free from the problem of finite-size effect,
and therefore one can choose the realistic magnitude of superconducting pair potential.
For $\gamma$ band, we assume two-dimensional chiral $p$-wave pair potential
for all the cases we have studied.
For $\alpha$ and $\beta$ bands,
we study two kinds of pair potentials:
two-dimensional pair potentials and quasi-one-dimensional ones.
In two-dimensional model, the calculated SDOS ($\sigma_s$) shows a
zero energy (zero bias) dip.
In the presence of a spin-orbit interaction (SOI), from atomic origins,
a small zero energy (zero bias) peak inside dip-like structure in SDOS ($\sigma_s$)
appears for the two-dimensional model.
For the quasi-one-dimensional model,
where the pair potential from $\gamma$-band is dominant,
the obtained SDOS ($\sigma_s$)
shows a zero energy (zero bias) peak in the absence of SOI.
On the other hand, this zero energy (zero bias) peak of the SDOS($\sigma_s$)
is suppressed by the SOI.
In the case of quasi-one-dimensional model where the pair potentials from $\alpha$ and $\beta$-bands are dominant,
the resulting $\sigma_s$ shows a two-step zero bias peak.
The experimentally obtained two-step structure with sharp ZBCP can be
explained by this quasi-one-dimensional model.

The organization of this paper is as follows.
In Sec. \ref{sec2} we discuss the general formulation of
the recursive Green's function.
In Sec. \ref{sec3}, we show the calculated results
of SDOS and $\sigma_{s}$ by the recursive Green's function.
We also show Andreev bound states obtained in finite system
to understand SDOS and $\sigma_{s}$ in detail.
Sec. \ref{sec4} provides the summary of our paper.

\section{Model and Formulations}\label{sec2}
In this section, we introduce a three-band model for Sr$_2$RuO$_4$
and briefly review the recursive Green's function method combined with M$\ddot{\rm o}$bius transformation proposed by Umerski.\cite{Umerski}
\subsection{Three-band model for Sr$_2$RuO$_4$}
In Sr$_2$RuO$_4$,
three cylindrical Fermi surface called $\alpha$-, $\beta$- and $\gamma$-bands
are obtained by the first principle calculations or ARPES measurements.
A tight-binding model considering the $d_{xy}$-, $d_{yz}$- and $d_{zx}$-orbitals
can describe these band structures,
\begin{equation}
\mathcal{H}=\mathcal{H}_{\rm kin}+\mathcal{H}_{\rm soi}+\mathcal{H}_{\rm pair}.
\end{equation}
The first term expresses the kinetic energy,
\begin{equation}
\mathcal{H}_{\rm kin}=\sum_{{\bm k},\sigma}\hat c^\dag_{{\bm k}\sigma}
\begin{pmatrix}
\varepsilon_{yz}({\bm k})&g({\bm k})&0\\
g({\bm k})&\varepsilon_{zx}({\bm k})&0\\
0&0&\varepsilon_{xy}({\bm k})
\end{pmatrix}c_{{\bm k}\sigma},\label{eq:kin}
\end{equation}
where, $\hat c_{{\bm k}\sigma}=(c_{{\bm k},\sigma}^{yz}, c_{{\bm k},\sigma}^{zx}, c_{{\bm k},-\sigma}^{xy})^T$ is the annihilation operators with momentum ${\bm k}$ and spin $\sigma$ (-$\sigma$) for $yz$- and $zx$-orbitals ($xy$-orbital).
Considering the hopping integral up to next nearest neighbor sites, the components of Eq. (\ref{eq:kin}) are given by,
\begin{align}
\varepsilon_{xy}({\bm k})&=-2t_1(\cos k_x+\cos k_y)-4t_2\cos k_x\cos k_y-\mu_{xy},\\
\varepsilon_{yz}({\bm k})&=-2t_4\cos k_x-2t_3\cos k_y-\mu_{yz},\\
\varepsilon_{zx}({\bm k})&=-2t_3\cos k_x-2t_4\cos k_y-\mu_{zx},\\
g({\bm k})&=-4t_5\sin k_x\sin k_y.
\end{align}
The second term describes an atomic SOI which causes a mixture of spin and orbital,
\begin{align}
\mathcal{H}_{\rm soi}&=\lambda\sum_{{\bm k},\sigma}
\hat c^\dag_{{\bm k}\sigma}
\begin{pmatrix}
0&is_\sigma&-s_\sigma\\
-is_\sigma&0&i\\
-s_\sigma&-i&0
\end{pmatrix}
\hat c_{{\bm k}\sigma},
\end{align}
where, $s_\sigma=1$ ($s_\sigma=-1$) for $\sigma=\uparrow$ ($\sigma=\downarrow$).
The last term denotes the condensation energy due to the formation of Cooper pairs.
In the superconducting state of Sr$_2$RuO$_4$, the most promising candidate of the pair potential belongs to the $E_u$ irreducible representation.\cite{Maeno2,Maeno2012}
In this irreducible representation, we consider the intraorbital pairing,
\begin{align}
\mathcal{H}_{\rm pair}=\sum_{{\bm k},\ell}\Delta^*_\ell({\bm k}) c^\ell_{{\bm k},\uparrow}c^\ell_{-{\bm k},\downarrow}+h.c.,
\end{align}
where, $\ell$ denotes the orbital index.
Considering the crystal symmetry of Sr$_2$RuO$_4$ and the orbital nature,
we study two kinds of pair potentials.
The first case is the two-dimensional pair potentials,
\begin{eqnarray}
\left\{
\begin{array}{l}
\Delta_{yz}({\bm k})=\Delta_{zx}({\bm k})=\Delta_{1}(\sin k_x+i\sin k_y),\\
\Delta_{xy}({\bm k})=\Delta_{2}(\sin k_x+i\sin k_y).
\end{array}
\right.\label{eq:pp2d}
\end{eqnarray}
Since the $d_{yz}$- and $d_{zx}$-orbitals have quasi-one-dimensional nature,
hence we choose another possibility which we call the quasi-one-dimensional pair potentials,
\begin{eqnarray}
\left\{
\begin{array}{l}
\Delta_{yz}({\bm k})=i\Delta_{1}\sin k_y,\\
\Delta_{zx}({\bm k})=\Delta_{1}\sin k_x,\\
\Delta_{xy}({\bm k})=\Delta_{2}(\sin k_x+i\sin k_y).
\end{array}
\right.\label{eq:pp1d}
\end{eqnarray}
\subsection{Recursive Green's function method}
Starting from the Hamiltonian introduced in the previous section,
we calculate local Green's function at the surface of semi-infinite Sr$_2$RuO$_4$ layer.
For this purpose, we use recursive Green's function method
using M$\ddot{\rm o}$bius transformation proposed by Umerski.\cite{Umerski}
For semi-infinite layer,
we consider the clean and homogeneous system with a flat (100) surface at $x=x_1$.
Then, we can assume that the momentum parallel to the surface, $k_y$,
is a good quantum number and the system is one-dimensional chain for each $k_y$.
To calculate the surface Green's function at $k_y$ and complex frequency $z$,
we first define the following matrix,
\begin{eqnarray}
X=
\begin{pmatrix}
\hat 0 &\hat t_{i+1,i}^{-1}(k_y)\\
-\hat t_{i,i+1}(k_y) & (z\hat I-\hat h_{loc}(k_y))\hat t_{i+1,i}^{-1}(k_y)
\end{pmatrix},
\end{eqnarray}
where, $\hat I$ is a unit matrix.
$\hat h_{loc}(k_y)$ and $\hat t_{i,j}(k_y)$ are matrices of local term and non-local term in the Hamiltonian, respectively.
The size of the Hamiltonian is $12\times12$ including
three, two and two degrees of freedom for orbital, spin and electron-hole spaces, respectively,
i.e., the size of the matrix $X$ is $24\times24$.
After diagonalization of $X$ by the eigenmatrix $O$,
\begin{eqnarray}
O^{-1}XO=\begin{pmatrix}
\begin{array}{cc}
\lambda_1&\\
&\lambda_2
\end{array}
&0\\
0&
\begin{array}{cc}
\ddots&\\
&\lambda_{24}
\end{array}
\end{pmatrix},
\end{eqnarray}
with $|\lambda_1|<|\lambda_2|<\cdots<|\lambda_{24}|$,
we obtain the surface Green's function $\hat G_{1s}(k_y, z)$ in the following form,
\begin{eqnarray}
\hat G_{1s}(k_y, z)&=&O_{12}O_{22}^{-1},
\end{eqnarray}
where $O_{12}$ and $O_{22}$ are the submatrices of $O$
\begin{eqnarray}
O&=&\begin{pmatrix}
O_{11}&O_{12}\\
O_{21}&O_{22}
\end{pmatrix}.
\end{eqnarray}
From the poles of this surface Green's function inside the energy gap,
we can evaluate the dispersion of the ABSs.
We can also evaluate the SDOS by the imaginary part of the retarded Green's function,
\begin{eqnarray}
\rho(\omega)=-\frac{1}{2\pi^2}\int^\pi_{-\pi}{\rm Im}\{{\rm Tr}' [\hat G_{1s}(k_y, \omega+i\eta)]\}dk_y,\label{eq:dos}
\end{eqnarray}
where $\eta$ is an infinitesimal imaginary part.
For the trace in Eq. (\ref{eq:dos}),
we only sum up the electronic part and drop the contribution due to the holes.
Next, we calculate the conductance in normal metal/superconducting Sr$_2$RuO$_4$ junction.
In the normal metal, we consider a two-dimensional single-band model with the energy dispersion given by
$\varepsilon({\bm k})=-2t_n(\cos k_x+\cos k_y)-4t_n'\cos k_x\cos k_y-\mu_n$,
where $t_n$ ($t_n'$) is the transfer integral between (next) nearest neighbor sites.
To calculate the conductance,
we first obtain the surface Green's functions
$\hat G_{0s}(k_y, z)$ for normal metal at $x=x_0$
and $\hat G_{1s}(k_y, z)$ for Sr$_2$RuO$_4$ at $x=x_1$
in the absence of the interface hoppings, and
we obtain the local and the non-local Green's functions
in normal metal/superconducting Sr$_2$RuO$_4$ junction,
\begin{eqnarray}
\hat G_{00}(k_y, z)&=&\{\hat G_{0s}(k_y, z)^{-1}-\hat t_{01}\hat G_{1s}(k_y, z)\hat t_{10}\}^{-1},\\
\hat G_{11}(k_y, z)&=&\{\hat G_{1s}(k_y, z)^{-1}-\hat t_{10}\hat G_{0s}(k_y, z)\hat t_{01}\}^{-1},\\
\hat G_{01}(k_y, z)&=&\hat G_{0s}(k_y, z)\hat t_{01}\hat G_{11}(k_y, z),\\
\hat G_{10}(k_y, z)&=&\hat G_{1s}(k_y, z)\hat t_{10}\hat G_{00}(k_y, z),
\end{eqnarray}
where, $\hat t_{01}$ and $\hat t_{10}$ are the hopping matrices at the interface.
From these Green's functions,
we obtain the conductance $\sigma_S$ by using the Kubo formula.\cite{Takane,Lee-Fisher}
Since we start from the BdG Hamiltonian,
which is not the eigenstate of the particle number,
for Sr$_2$RuO$_4$,
the electric current is not conserved inside the superconductor
unless either a source term is added
or self-consistency due to proximity effect is included.\cite{Josephson}
To avoid this problem, in the actual calculation,
we calculate the expectation values of the current operator inside the normal metal,
where the source term is absent.
Note that known single band results in the absence of SOI\cite{Sengupta} are reproduced
in the present three band model
if we just introduce the interface hopping
between normal metal and d$_{xy}$-orbital in Sr$_2$RuO$_4$.

\section{Results}\label{sec3}
\begin{figure}[htbp]
\begin{center}
\includegraphics[width=5cm]{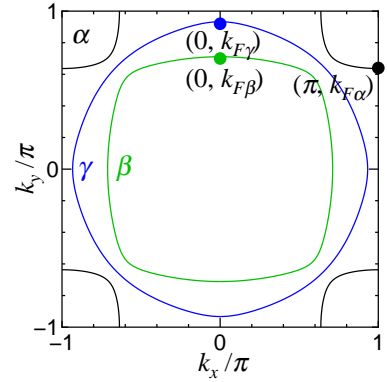}
\caption{(Color online) Fermi surfaces for $t_1=1$, $t_2=0.395$, $t_3=1.25$, $t_4=0.125$, $t_5=0.15$ and $\lambda=0$.
These for $\lambda=0.3$ are almost the same.}\label{fig3}
\end{center}
\end{figure}

In this section, we show the calculated results of the dispersion of ABSs, SDOS and conductance.
Prior to that, we explain the choice of parameters.
We have set $t_1=1$ as a unit of energy,
and for other hopping parameters in Sr$_2$RuO$_4$,
we choose $t_2=0.395$, $t_3=1.25$, $t_4=0.125$ and $t_5=0.15$
as adopted in the previous studies.\cite{Nomura05,Nomura08}
The obtained Fermi surfaces using these parameters are shown in Fig. \ref{fig3},
in which the Fermi momenta in $k_y$-direction
$k_{F\alpha}$, $k_{F\beta}$ and $k_{F\gamma}$
are defined for convenience.
For the SOI $\lambda$, we adopt the value estimated from the quasiparticle spectra of angle resolved photoemission
spectroscopy {\it et al}.\cite{Zabolotnyy}
The obtained values of $\lambda$ in Ref. \citen{Zabolotnyy} is $\lambda/t_1\simeq0.40$ and $\lambda/t_3\simeq0.22$,
which correspond to $\lambda\simeq0.40$ and $0.28$, respectively, by using the values of $t_1$ and $t_3$ in the above.
Though $\lambda\simeq0.3$ is the plausible value by the above estimation, we also show the results of ABSs and conductance
for $\lambda=0$, $0.1$ and $0.2$ to clarify the effect of the SOI.
The chemical potentials are determined to set the number of electron to be $n_{xy}=n_{yz}=n_{zx}=2/3$ per spin.
For the magnitude of the pair potential we choose $\Delta_0=0.001$ which is comparable to the critical temperature of Sr$_2$RuO$_4$.
For the hopping parameters and the chemical potential in the normal metal,
we choose $t_n=1$, $t_n'=0.395$ and $\mu_n=1.5005$,
which are the same as those for $xy$-orbital without the SOI ($\gamma$-band).
In the later subsections, we present the results for following three cases:
The first one is the two-dimensional pair potential as shown in Eq.(\ref{eq:pp2d}) with $\Delta_1=0.4\Delta_0$ and $\Delta_2=\Delta_0$.
This is a simple expansion of a single band model considering $\gamma$-band.\cite{Miyake,Nomura00,Ogata}
The second is the quasi-one-dimensional pair potential as shown in Eq.(\ref{eq:pp1d}) with $\Delta_1=0.4\Delta_0$ and $\Delta_2=\Delta_0$.
This pair potential includes quasi-one-dimensional nature of $\alpha$ and $\beta$-bands,
while the dominant pairing is composed of $\gamma$-band.\cite{Nomura02a,Nomura02b,Nomura05}
Finally, the quasi-one-dimensional pair potential is same as the second one
but with $\Delta_1=\Delta_0$ and $\Delta_2=0.4\Delta_0$.
In this case, the dominant pairing comes from $\alpha$ and $\beta$-bands.\cite{Raghu,Imai2012,Imai2013}

\subsection{Two-dimensional pair potential}
In this subsection, we consider the case of two-dimensional pair potential,
where the pair potential of each orbital is shown in Eq. (\ref{eq:pp2d}).
For the magnitude of the pair potential, we choose $\Delta_1=0.4\Delta_0$ and $\Delta_2=\Delta_0$
in accordance with theoretical results by Nomura and Yamada.\cite{Nomura05,Nomura08}
The interface hoppings in normal metal/Sr$_2$RuO$_4$ junction are chosen as $t_{xy}=t_{yz}=t_{zx}=1$ ($t_{xy}=t_{yz}=t_{zx}=0.2$)
for high (low) transmissivity.

\begin{figure}[htbp]
\begin{center}
\includegraphics[width=\columnwidth]{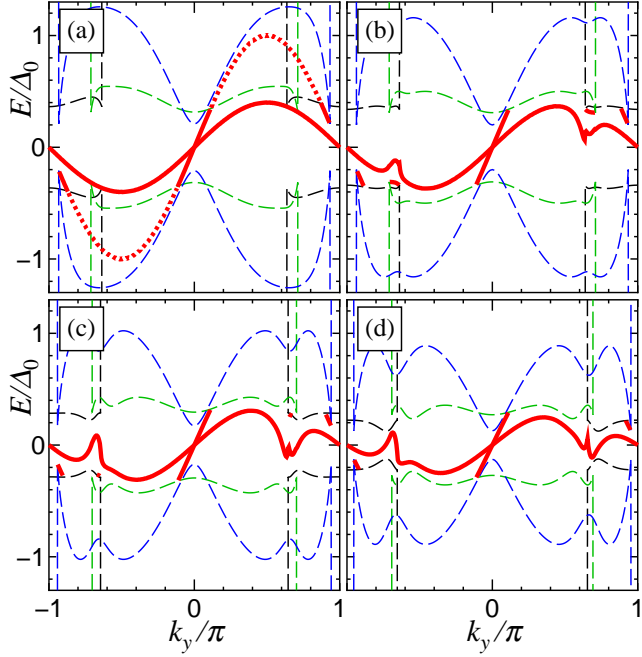}
\caption{(Color online) Energy spectrum for two-dimensional pair potential with $\Delta_1=0.4\Delta_0$ and $\Delta_2=\Delta_0$
for (a) $\lambda=0$, (b) $\lambda=0.1$, (c) $\lambda=0.2$ and (d) $\lambda=0.3$.
Thin broken lines show the bulk energy gap on the Fermi surfaces.
Thick solid lines show the dispersion of ABS inside the bulk energy gap.
Thick dotted lines show the ABS in the continuum levels ($\lambda=0$).
}\label{fig4}
\end{center}
\end{figure}

\begin{figure}[htbp]
\begin{center}
\includegraphics[width=6cm]{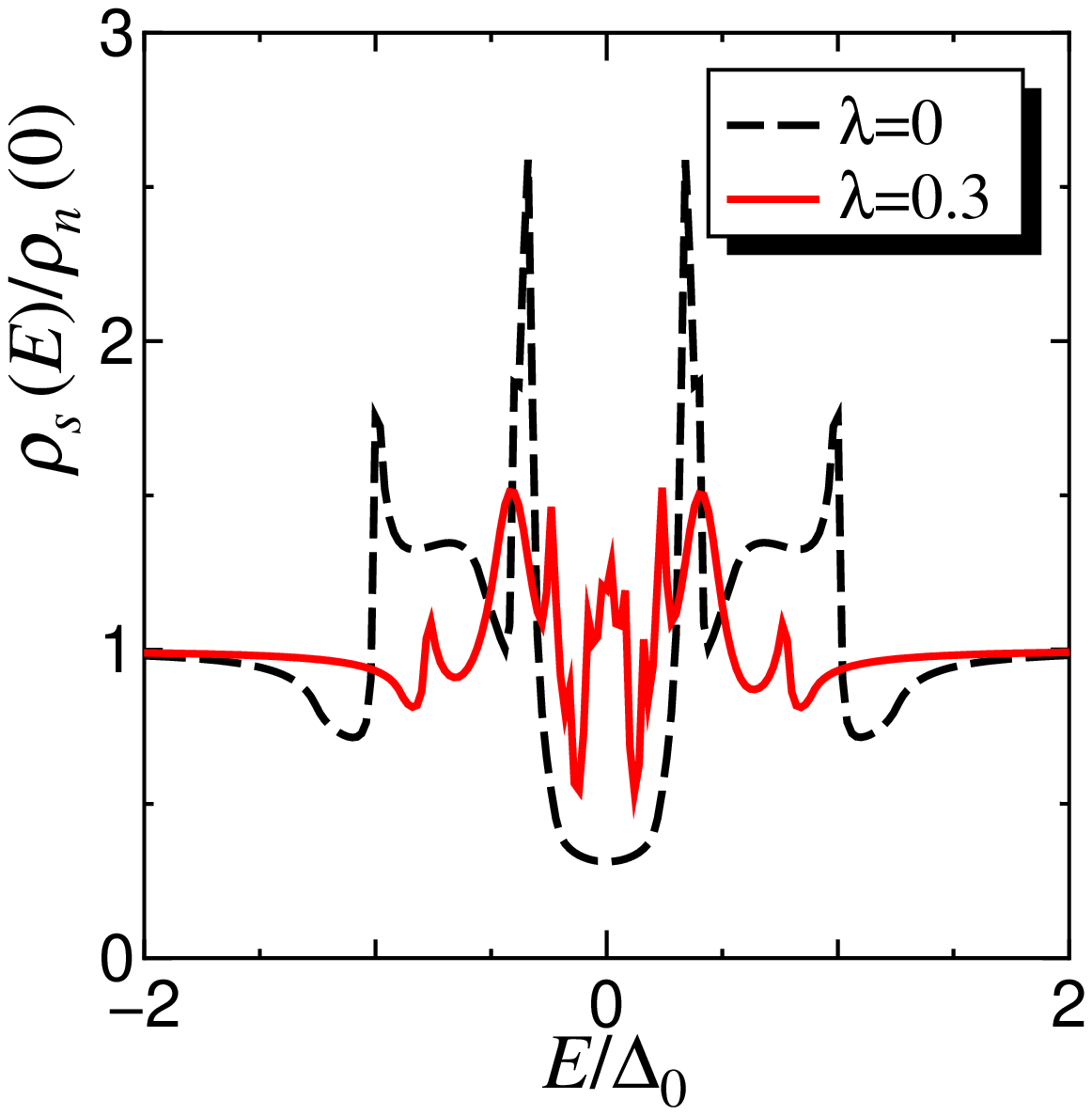}
\caption{
(Color online) Normalized SDOS for two-dimensional pair potential with $\Delta_1=0.4\Delta_0$ and $\Delta_2=\Delta_0$
for $\lambda=0$ (dashed line) and $\lambda=0.3$ (solid line).
}\label{fig5}
\end{center}
\end{figure}

\begin{figure}[htbp]
\begin{center}
\includegraphics[width=8.5cm]{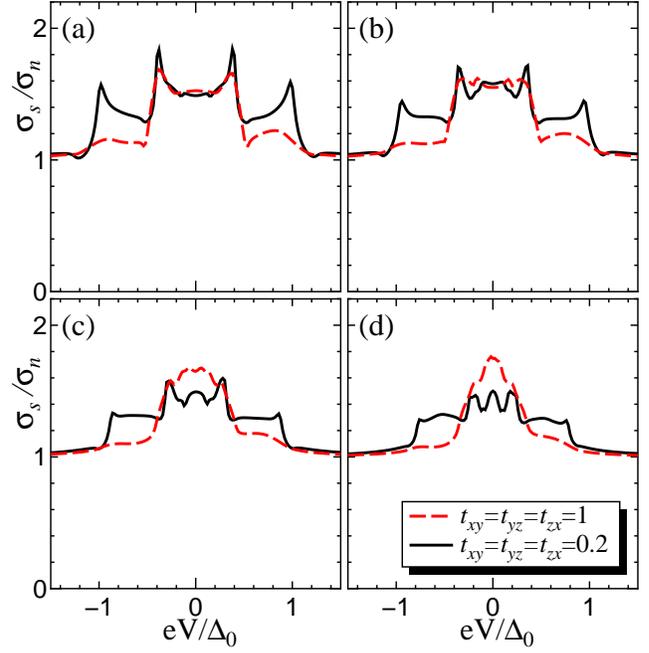}
\caption{
(Color online) Normalized conductance for two-dimensional pair potential with $\Delta_1=0.4\Delta_0$ and $\Delta_2=\Delta_0$
for (a) $\lambda=0$, (b) $\lambda=0.1$, (c) $\lambda=0.2$ and (d) $\lambda=0.3$.
Interface hoppings are chosen as $t_{xy}=t_{yz}=t_{zx}=1$ (dashed line) and $t_{xy}=t_{yz}=t_{zx}=0.2$ (solid line) for high and low transmissivity, respectively.
}\label{fig6}
\end{center}
\end{figure}

The calculated dispersion of ABSs is shown in Fig. \ref{fig4}.
Since Sr$_2$RuO$_4$ has a mirror symmetry, the Hamiltonian of Sr$_2$RuO$_4$ is divided into two mirror sectors.\cite{Ueno}
In Fig. \ref{fig4}, we show the ABSs for one of the mirror sectors.
The ABSs for the other sector is obtained by the particle-hole transformation, i.e. $E(k_y)\rightarrow -E(-k_y)$.
In the absence of SOI, there are three ABS originating from $\alpha$, $\beta$, and $\gamma$-bands as shown in Fig. \ref{fig4} (a).
Dispersions of these three ABS is represented by $\Delta_1\sin k_y$ ($\Delta_2\sin k_y$) for $\alpha$ and $\beta$-bands ($\gamma$-band).
Note that the ABSs that originate from $\alpha$ and $\beta$-bands are degenerate
at $k_{F\alpha}<|k_y|<k_{F\beta}$;
where, $k_{F\alpha}$ and $k_{F\beta}$ are the Fermi momenta shown in Fig. \ref{fig3}.
This degeneracy is lifted by SOI and one of the ABS disappears at $\lambda=0.3$ as shown in Figs. \ref{fig4} (b)-(d).
In these lifted ABSs, the one with the higher-energy
disappears as increasing the magnitude of SOI.
The lower-energy one is strongly suppressed and it crosses the Fermi level for $\lambda\ge0.2$.
Next, we show calculated results for
the SDOS and conductance in Figs. \ref{fig5} and \ref{fig6}, respectively.
Without the SOI,
SDOS shows a zero energy dip surrounded by four peaks
at around $E=\pm\Delta_1$ and $\pm\Delta_2$.
The position of the peaks corresponds to the energy levels of the ABSs
at around $k_y=\pm0.5\pi$ where their slope become to be zero.
These features are essentially the same as the single band results considering the pair potential $\Delta({\bm k})=\Delta_0(\sin k_x+i\sin k_y)$.\cite{Sengupta}
Further, $\sigma_s$ in low transmissivity has a zero bias dip
and four peaks at $eV=\pm\Delta_1$, $\pm\Delta_2$ similar to the SDOS
as seen from solid line in Fig. \ref{fig6}(a).
Though these four peaks are smeared in high transmissivity,
$\sigma_s$ still has a dip-like structure at $eV\sim0$.
On the other hand, SDOS in the presence of SOI has zero bias peak
since the diepersion of ABS at $k_{F\alpha}<|k_y|<k_{F\beta}$ is close to the Fermi levels.
It also have many small spikes reflecting the complex structure of the energy dispersion of the ABS.
$\sigma_s$ in the presence of SOI has a zero bias peak for both high and low transmissivity as seen from Fig. \ref{fig6}(b)-(d).
In the case of low transmissivity, as compared to $\sigma_s$ without SOI,
the positions of peaks in $\sigma_s$ with SOI other than zero bias peak shift to lower energy and the height of the peaks becomes lower
due to the suppression of the bulk energy gap.
In the case of high transmissivity, three peaks merge and $\sigma_s$ shows a single ZBCP.

\subsection{Quasi-one-dimensional pair potential}
Here, we consider the case of quasi-one-dimensional pair potential
where the pair potential of each orbital is given by Eq. (\ref{eq:pp1d}).
We consider two choices for the magnitude of the pair potentials.
One is the case with $\Delta_1=0.4\Delta_0$, $\Delta_2=\Delta_0$.
This pair potential well reproduces Nomura and Yamada's results\cite{Nomura05,Nomura08}
not only for the magnitude of pair potential but for the momentum dependence of the pair potential on the Fermi surface.
The other case is $\Delta_1=\Delta_0$, $\Delta_2=0.4\Delta_0$ to emphasize the importance of the quasi-one-dimensional band
as suggested by some microscopic models.\cite{Raghu,Imai2012,Imai2013}

\begin{figure}[htbp]
\begin{center}
\includegraphics[width=\columnwidth]{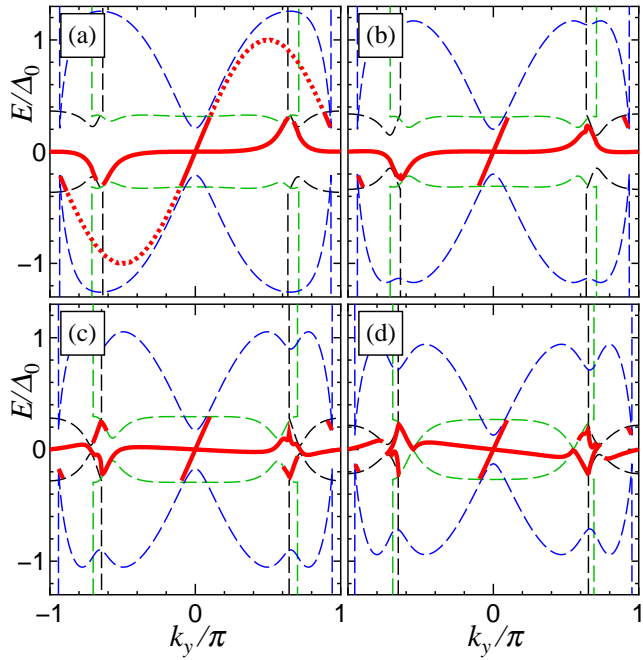}
\caption{(Color online) Energy spectrum for quasi-one-dimensional pair potential with $\Delta_1=0.4\Delta_0$ and $\Delta_2=\Delta_0$
for (a) $\lambda=0$, (b) $\lambda=0.1$, (c) $\lambda=0.2$ and (d) $\lambda=0.3$.
Thin broken lines show the bulk energy gap on the Fermi surfaces.
Thick solid lines show the dispersion of ABS inside the bulk energy gap.
Thick dotted lines show the ABS in the continuum levels ($\lambda=0$).}\label{fig7}
\end{center}
\end{figure}

\begin{figure}[htbp]
\begin{center}
\includegraphics[width=6cm]{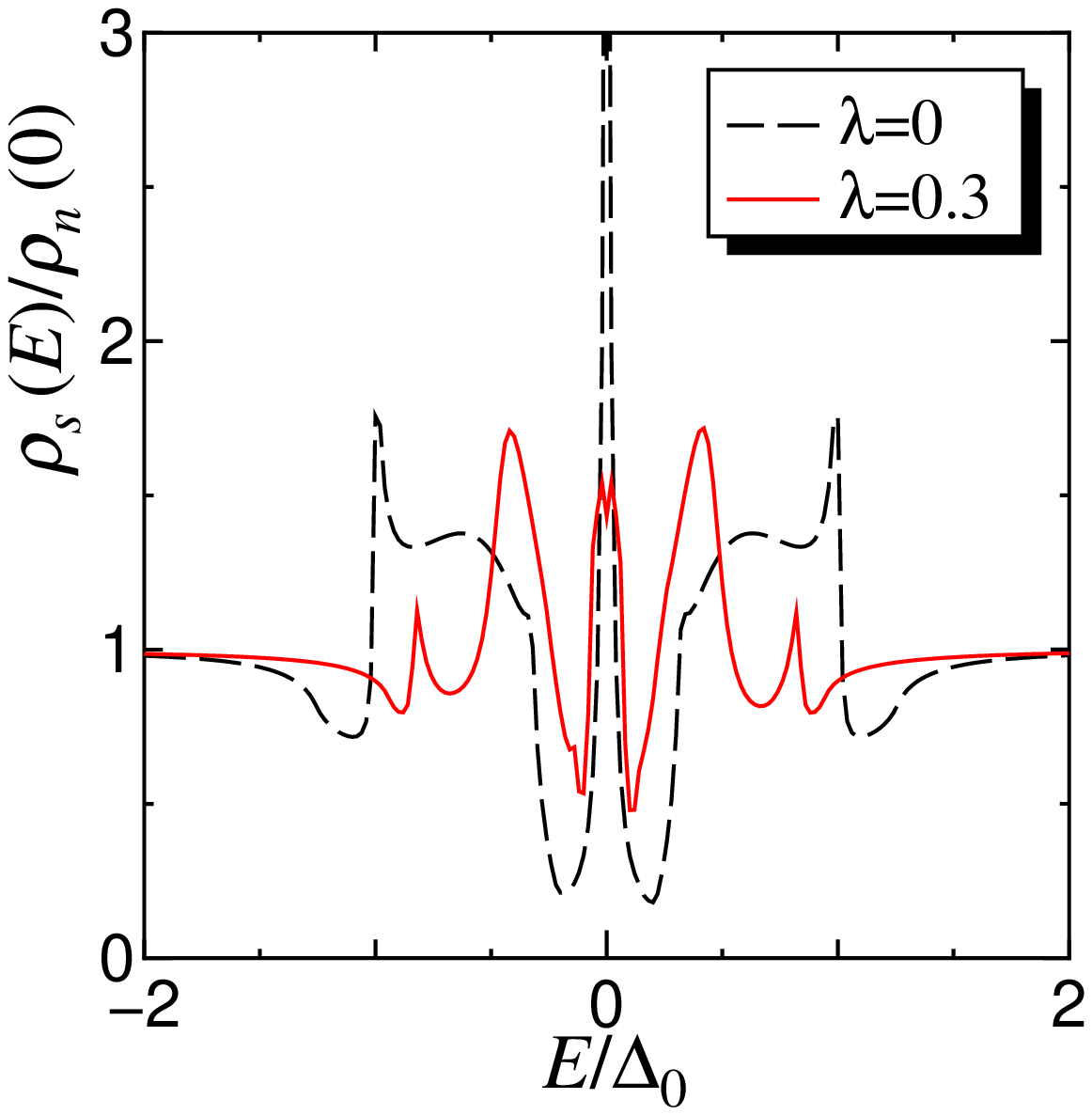}
\caption{
(Color online) Normalized SDOS for quasi-one-dimensional pair potential with $\Delta_1=0.4\Delta_0$ and $\Delta_2=\Delta_0$
for $\lambda=0$ (dashed line) and $\lambda=0.3$ (solid line).
}\label{fig8}
\end{center}
\end{figure}

\begin{figure}[htbp]
\begin{center}
\includegraphics[width=8.5cm]{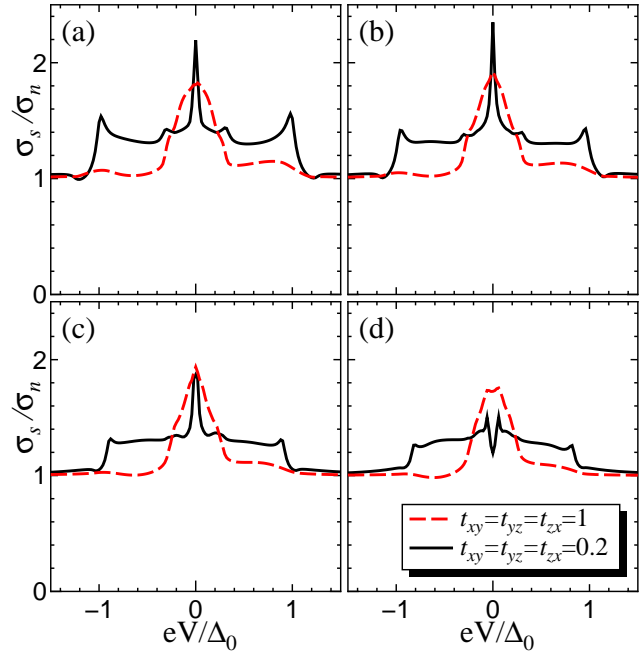}
\caption{
(Color online) Normalized conductance for quasi-one-dimensional pair potential with $\Delta_1=0.4\Delta_0$ and $\Delta_2=\Delta_0$
for (a) $\lambda=0$, (b) $\lambda=0.1$, (c) $\lambda=0.2$ and (d) $\lambda=0.3$.
Interface hoppings are chosen as $t_{xy}=t_{yz}=t_{zx}=1$ (dashed line) and $t_{xy}=t_{yz}=t_{zx}=0.2$ (solid line) for high and low transmissivity, respectively.
}\label{fig9}
\end{center}
\end{figure}

First, we consider the case for $\Delta_1=0.4\Delta_0$ and $\Delta_2=\Delta_0$
without the SOI.
Without SOI, bulk energy gap and dispersion of ABS (\ref{fig7}(a))
originating from $\gamma$-band
are the same as in the case of two-dimensional pair potential (Fig. \ref{fig4}(a))
because the pair potential for $xy$-orbital are the same.
On the other hand, the dispersion of chiral edge modes originating from $\alpha$ and $\beta$-bands are almost flat at around $k_y=0$ and $\pm\pi$.
This can be understood as follows:
On the Fermi surface of $\alpha$- and $\beta$-bands,
where azimuthal angles measured from the center of the Fermi surfaces are from $-\pi/4$ to $\pi/4$, i.e. at around $k_y=0$ ($\beta$-band) and $\pm\pi$ ($\alpha$-band),
their dominant components of orbitals are $zx$-orbital,
whose pair potential is proportional to $\sin k_x$ like $p_x$-wave.
It is known that the system has zero energy ABS on (100) surface for $p_x$-wave pair potential
due to the $\pi$-phase shift of the pair potential between incident and reflected quasi-particles.\cite{Tanuma2001,ABSb,DasSarma,Buchholtz}
This one-dimensional nature remains even in the presence of the mixture of $yz$ and $zx$-orbitals.
For this reason, the energy levels of the ABS near $k_y=0$ and $\pm\pi$ is nearly zero.
With increasing the magnitude of the SOI,
quasi-one-dimensional nature of the ABS disappears due to the coupling with $xy$-orbital as seen from Figs. \ref{fig7}(b)-(d).
Due to the level repulsion between the ABSs originating from $\gamma$-band and $\beta$-bands,
the group velocity of the ABS of $\beta$-bands at $k_y\sim0$ becomes negative.
The resulting SDOS and conductance are shown in Fig. \ref{fig8} and Fig. \ref{fig9}, respectively.
SDOS at low energy and $\sigma_s$ in low bias voltage (e.g. $|E/\Delta_0|<0.5$ and $|eV/\Delta_0|<0.5$, respectively)
drastically change as compared to those with two-dimensional pair potential.
SDOS in the absence of the SOI shows a sharp zero energy peak
due to the quasi-one-dimensional nature of the ABS as seen from dashed line in Fig. \ref{fig8}.
A similar line shape is also found in $\sigma_s$ in the low transmissivity for $\lambda$ up to 0.2(Fig. \ref{fig9}(a)-(c)).
In contrast, this sharp zero energy (bias) peak in SDOS ($\sigma_s$ in low transmissivity) suppressed for $\lambda=0.3$,
and it shows a small zero bias dip as shown in Fig. \ref{fig8} (Fig. \ref{fig9}(d)).
In high transmissivity, $\sigma_s$ shows a single zero bias peak regardless of the magnitude of the SOI.

\begin{figure}[htbp]
\begin{center}
\includegraphics[width=\columnwidth]{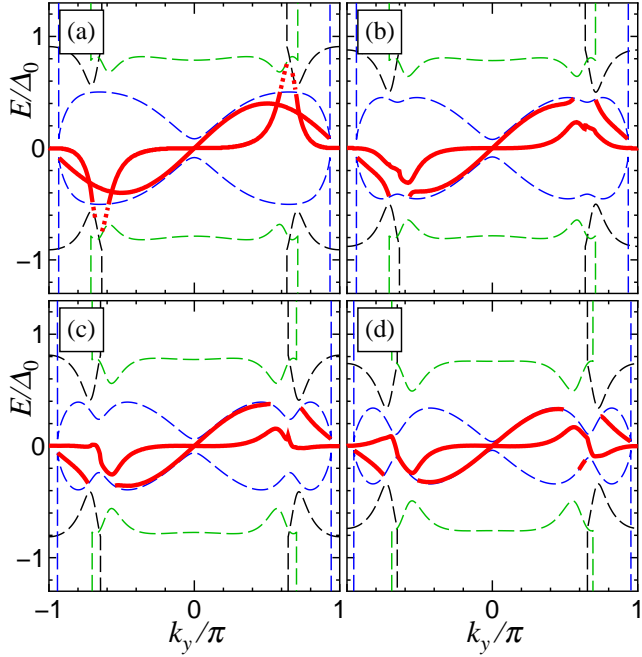}
\caption{(Color online) Energy spectrum for quasi-one-dimensional pair potential with $\Delta_1=\Delta_0$ and $\Delta_2=0.4\Delta_0$
for (a) $\lambda=0$, (b) $\lambda=0.1$, (c) $\lambda=0.2$ and (d) $\lambda=0.3$.
Thin broken lines show the bulk energy gap on the Fermi surfaces.
Thick solid lines show the dispersion of ABS inside the bulk energy gap.
Thick dotted lines show the ABS in the continuum levels ($\lambda=0$).
}\label{fig10}
\end{center}
\end{figure}

\begin{figure}[htbp]
\begin{center}
\includegraphics[width=6cm]{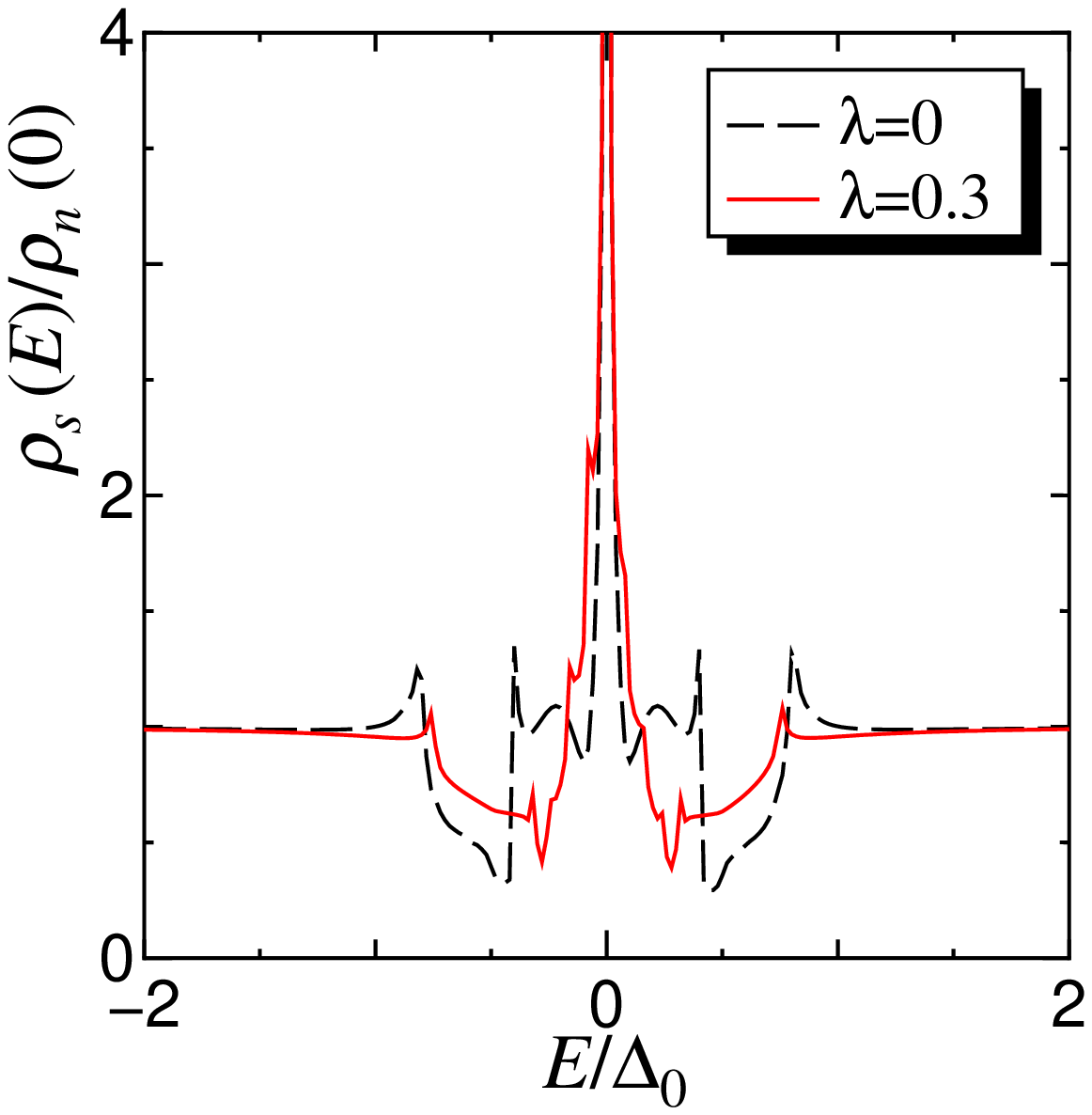}
\caption{(Color online) Normalized SDOS for quasi-one-dimensional pair potential with $\Delta_1=\Delta_0$ and $\Delta_2=0.4\Delta_0$
for $\lambda=0$ (dashed line) and $\lambda=0.3$ (solid line).
}\label{fig11}
\end{center}
\end{figure}

\begin{figure}[htbp]
\begin{center}
\includegraphics[width=8.5cm]{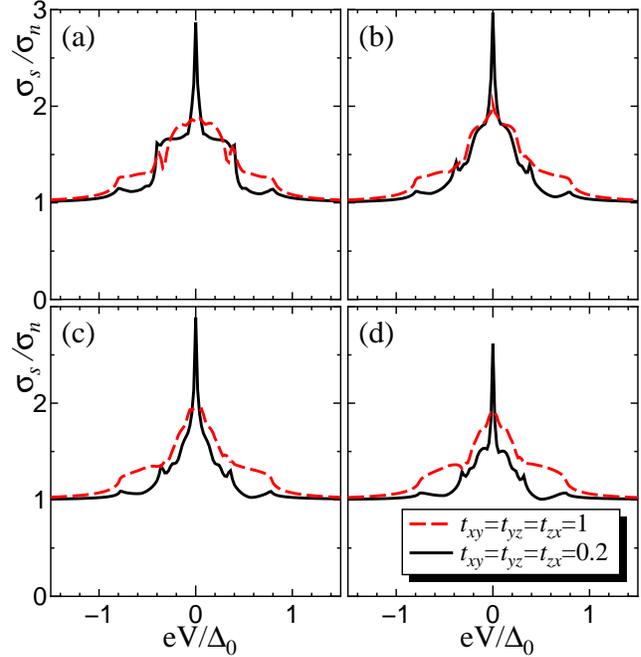}
\caption{
(Color online) Normalized conductance for quasi-one-dimensional pair potential with $\Delta_1=\Delta_0$ and $\Delta_2=0.4\Delta_0$
for (a) $\lambda=0$, (b) $\lambda=0.1$, (c) $\lambda=0.2$ and (d) $\lambda=0.3$.
Interface hoppings are chosen as $t_{xy}=t_{yz}=t_{zx}=1$ (dashed line) and $t_{xy}=t_{yz}=t_{zx}=0.2$ (solid line) for high and low transmissivity, respectively.
}\label{fig12}
\end{center}
\end{figure}

Next, we consider another choice of the magnitude of the pair potential;
$\Delta_1=\Delta_0$ and $\Delta_2=0.4\Delta_0$.
The dispersions of the ABSs without the SOI shown in Fig. \ref{fig10}(a) are identical with those for the previous case (Fig. \ref{fig7}(a))
except for the energy scales of $\Delta_1$ and $\Delta_2$.
Thus, the SDOS and $\sigma_s$ show a sharp zero energy peak and a ZBCP similar to the case with $\Delta_2>\Delta_1$
as shown in Fig. \ref{fig11} and Fig. \ref{fig12}(a), respectively.
The dispersions of the ABSs in the presence of the SOI are shown in Figs. \ref{fig10}(b)-(d).
In comparison with the case of $\Delta_1<\Delta_2$ in Fig. \ref{fig11},
the effect of the SOI on the group velocity of the ABSs at around $k_y=0$ and $\pm\pi$ is small.
This is because the induced two-dimensionality for the ABS originating from $\alpha$- $\beta$-bands is small
since the magnitude of the two-dimensional pair potential in $xy$-orbital is small.
As a result, the SDOS and $\sigma_s$ still have zero energy peak and ZBCP up to $\lambda=0.3$
as shown in Fig. \ref{fig11} and Fig. \ref{fig12}(b)-(d), respectively.
Note that, we can not see clear dip-like structure in $\sigma_s$
originating from $\gamma$-band under cover of strong ZBCP due to $\alpha$- $\beta$-bands.
Thus, $\sigma_s$ shows the two-step peaks regardless of the magnitude of the SOI
as shown in solid lines in Fig. \ref{fig12}.

\section{Discussion and Summary}\label{sec4}
\begin{table}[htbp]
\begin{center}
\begin{tabular}{|c|c|c|}
\hline
pair potential& dominant pair &line shape of conductance\\
\hline
2D($\alpha,\beta,\gamma$)& $\gamma$ & Three peaks \\
Q1D($\alpha,\beta$)+2D($\gamma$)& $\gamma$ & Tiny zero bias dip \\
Q1D($\alpha,\beta$)+2D($\gamma$)& $\alpha,\beta$ & Two-step peak \\
$k_x+ik_y$& - & single ZBCP\\
$\sin k_x+i\sin k_y$& - & Zero bias dip\\
\hline
\end{tabular}
\caption{Line shapes of the conductance for low transmissivity in the present multi-band model for $\lambda=0.3$ (upper three rows) and in single-band models (lower two rows).}\label{table1}
\end{center}
\end{table}
In table \ref{table1}, we show our results of the line shapes of conductance for $\lambda=0.3$ in the three-band model
as well as those in a single-band model in the previous studies\cite{YTK97,YTK98,Sengupta}.
Here, we compare the calculated results with experimental data shown in Fig. \ref{fig1}.
Two-step ZBCP structure like curve (c) in Fig. \ref{fig1}
only appears for quasi-one-dimensional pair potential with $|\Delta_{1}|>|\Delta_{2}|$ (see Fig. \ref{fig12}).
We shall discuss the strength of the SOI here.
In Ref. \citen{Haverkort}, the band dispersion of Sr$_2$RuO$_4$ has been studied by the first principle calculation including the SOI.
The obtained value of the energy-level splitting at $\Gamma$-point between $\alpha$ and $\beta$-bands is about 90meV.
In the present model, by choosing the unit of the energy $t_1$ to be 230meV,
we can reproduce the energy level of the $\alpha$ and $\beta$-bands at $\Gamma$-point.
The obtained values of the energy-level splitting are $48$, $98$ and $152$ meV for $\lambda=0.1$, 0.2 and 0.3, respectively.
If we simply see this result, $\lambda\sim0.2$ is the appropriate choice of the SOI.
In this case, two-step ZBCP also appears for the quasi-one-dimensional pair potential with $|\Delta_{1}|<|\Delta_{2}|$ as shown in Fig. \ref{fig9}(c).
However, the electron correlation which is not fully taken into account in the local density approximation might induce the renormalization of the quasiparticle energy bands
and the effective values of $\lambda/t_1$ becomes about $0.3$ as mentioned in Sec. \ref{sec3}.
In either case of $\lambda/t_1=0.2$ or $0.3$, two-step ZBCP never appears for the two-dimensional pair potential but for the quasi-one-dimensional pair potential.

As for a broad ZBCP like curve (b) in Fig. \ref{fig1}, a junction with high transmissivity shows this for all pairings considered in the present paper
since the Andreev reflection process governs the conductance in the energy gap.
A single ZBCP also appears when the size of the Fermi surface is small and/or the insulating barrier potential is high.
In these cases, the contribution of the perpendicular injection ($k_y=0$) is enhanced,
where the energy levels of the ABSs are close to zero for all the pairing considered in the present study.
However, the former is not likely in the present case since it is known that the size of the Fermi surface of Gold is large.

Besides, we have also confirmed that dip-like structure as shown in curve (a) in Fig. \ref{fig1} is reproduced
if we ignore the interface hoppings to $d_{yz}$- and $d_{zx}$-orbitals,
though the mechanism to realize this situation is unclear.
Nevertheless, if only the $\gamma$-band contributes to the conductance,
dip-like structure appears as explained in Ref. \citen{Sengupta}.
Another possibility of the emergence of a zero bias dip is the $c$-axis tunneling,
where there is no ABS and a conventional gap structure appears.
However, the junctions which are used in the measurements of the conductance,
have been made at the position without the $c$-axis tunneling by measuring the interface by the scanning ion microscopy.
Therefore, the contribution of the $c$-axis tunneling is excluded below the resolution.


In this paper, we have studied
ABSs, SDOS and $\sigma_s$ of Sr$_{2}$RuO$_{4}$
based on the three band model using recursive Green's function.
For 
$d_{xy}$-orbital, we assume two-dimensional chiral $p$-wave pair potential
$\Delta_{2}(\sin k_{x} + i \sin k_{y})$ for all cases we have studied.
For $d_{yz}$ and $d_{zx}$-orbital
we have studied two kinds of pair potentials:
two-dimensional pair potentials and quasi-one-dimensional ones.
In the former case, we have chosen
$\Delta_{yz}({\bm k})=\Delta_{zx}({\bm k})=\Delta_{1}(\sin k_x+i\sin k_y)$,
and for the latter,
$\Delta_{yz}({\bm k})=i\Delta_{1}\sin k_{y}$
and $\Delta_{zx}({\bm k})=\Delta_{1}\sin k_{x}$.
For a two-dimensional model, the calculated SDOS ($\sigma_s$) shows a
zero energy (bias) dip without the SOI.
In the presence of the SOI,
small zero energy (bias) peak inside dip-like structure in SDOS($\sigma_s$) appears.
While, in the quasi-one-dimensional model with $|\Delta_1|<|\Delta_2|$,
the obtained SDOS ($\sigma_s$) shows a zero energy (bias) peak.
This zero energy (bias) peak in SDOS ($\sigma_s$)
is suppressed by the SOI.
In the case of $|\Delta_1|>|\Delta_2|$, the resulting $\sigma_s$ shows a two-step zero bias peak
with sharp ZBCP and broad one.
The last one can reasonably explain the experimental data.
We do not consider the roughness of the surface and disorder.
Since the pairing symmetry is influenced by these effects,
it is necessary to construct more realistic theory taking account of these effects in the future.

\begin{acknowledgment}
We gratefully acknowledge M. Sato, A. Yamakage, I. A. Devyatov, A. V. Burmistrova and S. Onari for valuable discussions,
and we thank A. Dutt for critical reading of the manuscript.
This work was supported in part by
a Grant-in Aid for Scientific Research from MEXT of Japan, "Topological Quantum Phenomena" Grants No. 22103005 and No. 20654030 (Y.T.),
Dutch Foundation for Fundamental Research on Matter (FOM) and by EU-Japan program "IRON SEA",
and Russian Ministry of Education and Science.
\end{acknowledgment}

\bibliographystyle{jpsj}
\bibliography{66543}

\end{document}